\documentclass[a4paper]{spie}
\usepackage{graphicx, cite, amsmath, amssymb, subfigure}
\DeclareMathOperator{\abs}{abs}

\title{Mutual information for examining correlations in DNA}
\author{Matthew J. Berryman\supit{a}, Andrew Allison\supit{a}, and Derek Abbott\supit{a}
\skiplinehalf
\supit{a}Centre for Biomedical Engineering and\\
School of Electrical and Electronic Engineering,\\
The University of Adelaide, SA  5005, Australia.}
\authorinfo{Send correspondence to Derek Abbott\\
E-mail: dabbott@eleceng.adelaide.edu.au, Telephone: +61 8 8303 5748}
\begin{document} 
\maketitle
\begin{abstract}
This paper examines two methods for finding whether long-range correlations exist in DNA: a fractal measure and a mutual information technique. We evaluate the performance and implications of these methods in detail. In particular we explore their use comparing DNA sequences from a variety of sources. Using software for performing {\it in silico} mutations, we also consider evolutionary events leading to long range correlations and analyse these correlations using the techniques presented. Comparisons are made between these virtual sequences, randomly generated sequences, and real sequences. We also explore correlations in chromosomes from different species.
\end{abstract}

\section{Introduction}
DNA is a structure containing a long sequence of complimentary pairing bases, denoted by the symbol set $\left\{a,t,c,g\right\}$~\cite{DrWatson}. The genetic material in DNA undergoes a variety of different mutational events~\cite{ProkaryoticGenetics,MrDarwin}. These mutational events can be considered as string rewriting rules~\cite{BSA} that lead to correlations in DNA. Repeated use of short sequences as promoters~\cite{LevineTjian}, or as intron markers~\cite{Introns} can give rise to very long-range correlations. 

A number of different techniques have been studied for examining long range correlations in DNA. These include L\'{e}vy walks~\cite{dnawalk1}, Fourier transforms~\cite{Li,GSP,GSP2}, and wavelets~\cite{wavelets2}.
A number of people have attempted to explore this by considering power law relationships in power spectra of DNA sequences. This purports to show long-range correlations and also to show differences between regions of DNA. In this paper we examine long-range correlations with mutual information techniques~\cite{mutualinfo}, and briefly explore the Higuchi fractal method~\cite{Higuchi}.

DNA sequences contain a number of coding regions. These are regions that code for protein and are marked with a stop and start codons (but the presence of these does not necessarily indicate a coding region). Coding regions may contain introns, which are regions that get spliced (cut) out before translation from the RNA template before the protein is made according to the code on the RNA template (which in turn comes from the DNA).
Non-coding regions may just be junk, or may code for regulatory RNAs~\cite{mattick}, such as the Xist gene which switches off the extra X chromosome in women~\cite{XIST2}.

In this paper we show that these long-range correlations exist for real sequences of DNA and virtual sequences of DNA, but not random sequences of DNA. The virtual sequences of DNA are those produced by our software, which simulates a variety of mutational events. The random DNA has a random sequence generated in software, so it should contain almost no correlations. We also explore whether or not the power spectra show any differences between coding and non-coding DNA, and between different species of bacteria.
\section{Sequences Examined}
For exploring correlations at very large distances, we used {\it Homo sapiens} chromosome 20~\cite{HS_CHR20}, {\it Mus musculus} chromosome 2~\cite{MM1,MM2} and {\it Escherechia coli}~\cite{EColi}.
\subsection{Real sequences}
In order to compare correlations in real DNA with those in short random and short virtual DNA sequences, we chose a selection of twenty short, real gene sequences from various organisms. Their accession numbers, and descriptions are shown in Table~\ref{real}.
\begin{table}[htbp]
\centering
\caption{These are the GenBank~\cite{GenBank} accession numbers and descriptions of the twenty short, real mRNA sequences used.}
\begin{tabular}{|l|l|}\hline
NM\_076575 & {\it C.~elegans} essential {\it Drosophila} huncback like. \\\hline
NM\_169234 & {\it Drosophila melanogaster} hunchback CG9786-PB (hb).\\\hline
BC016664 & {\it Homo sapiens} cone-rod homeobox.\\\hline
BC016502 & {\it Mus musculus} cone-rod homeobox containing gene.\\\hline
NM\_031888 & {\it Homo sapiens} pro-melanin-concentrating hormone-like 2. \\\hline
NM\_010410 & {\it Rattus norvegicus} $\beta$-catenin.\\\hline
AY438620 & {\it Arabidopsis thaliana} GLUR3 (At1g05200) mRNA.\\\hline
AY148346 & {\it Mus musculus} sentrin-specific protease.\\\hline
BC062048 & {\it Rattus norvegicus} MAP kinase-activated protein kinase 2.\\\hline
BC002377 & {\it Homo sapiens} PTK7 protein tyrosine kinase 7.\\\hline
NM\_001437 & {\it Homo sapiens} estrogen receptor 2 (ESR2). \\\hline
BC057647 & {\it Mus musculus} visual system homeobox 1 homolog.\\\hline
BC060890 & {\it Danio rerio} retinal homeobox gene 1.\\\hline
BC004108 & {\it Homo sapiens} immunoglobulin superfamily, member 8.\\\hline
BC048387 & {\it Mus musculus} immunoglobulin superfamily, member 8.\\\hline
NM\_033615 & {\it Mus musculus} ADAM33. \\\hline
NM\_025220 & {\it Homo sapiens} ADAM33, transcript variant 1. \\\hline
BC062067 & {\it Rattus norvegicus} SRY-box containing gene 10.\\\hline
BC002824 & {\it Homo sapiens} SRY-box containing gene 10. \\\hline
XM\_128139 & {\it Mus musculus} SRY-box containing gene 10.\\\hline
\end{tabular}
\label{real}
\end{table}
\subsection{Random sequences}
To compare the mutual information in real and virtual sequences, we generated twenty random sequences of length 10~000 bases, where all four bases have equal probability of appearing in each position.
\subsection{Virtual sequences} 
\begin{figure}[htbp]
\centering{\resizebox{7cm}{!}{\includegraphics{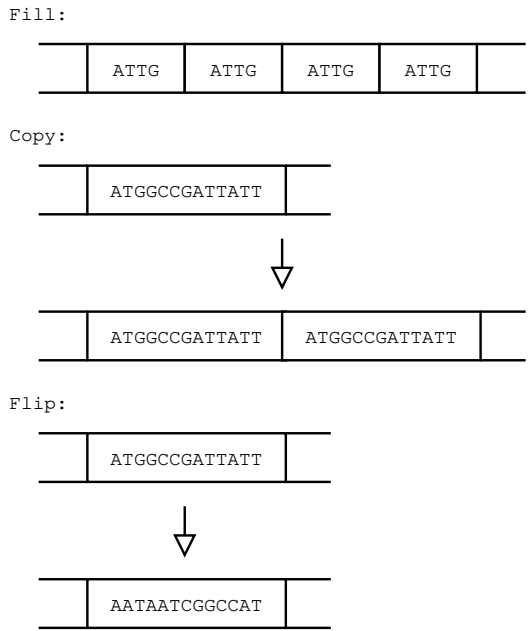}}}
  \caption{This figure shows the three operations:
  fill, where we have added a sequence of repetitive elements of length 4 in this case;
  copy, where we have copied part of a DNA sequence; and
  flip, where we have replaced part of the DNA sequence by its reverse complement.
  }
\label{mutations}
\end{figure}

The twenty virtual non-coding
regions are generated by the latest version of our software for exploring mutations
in DNA~\cite{MFC}. It implements the following {\it in silico} operations:
\begin{itemize}
\item Base substitutions, where one base pair has been replaced with a different base
through some mechanism (such as UV irradiation with an absent or partly
unsuccessful repair process).
\item Additions, where a base pair has been added to the sequence.
\item Deletions, where a base pair has been removed from the sequence.
\item Flips, where part of a sequence has been replaced by its reverse complement.
\item Fills, where a sequence of repetitive elements (of length 1 to 4) has been
inserted up to 50 times. The exact number of repetitions is chosen at random from a uniform distribution, as is the length.
\item Copies, where part of a sequence (up to 100 bases in length) has been copied. As with the fill operations, the length is chosen from a uniform random distribution.
\end{itemize}
The flip, fill, and copy operations are illustrated in Fig.~\ref{mutations}.
These operations are meant to simulate small scale general mutations, and larger
scale ones of the type that occur in non-coding DNA. In each run of the simulator
we took one of the random DNA sequences and used up to 30, the exact number chosen from a uniform random distribution, of each of the above mechanisms to generate long-range correlations in the DNA sequences. 
With some experimentation we found that, as
one would expect, the fill and copy mechanisms are the primary drivers in creating
long-range correlations.
\section{Methods for Exploring Correlations in DNA}
\subsection{Mutual information functions}
Another method for showing the existence of long-range correlations in DNA is to use the mutual information function, as given in Eq.~\ref{mutualinfo} below. This approach has been shown to distinguish between coding and non-coding regions~\cite{generatingcorr}.
We explore the use of the the mutual information function given in Eq.~\ref{mutualinfo}:
\begin{equation}
M(d)=\displaystyle \sum_{\alpha \in \mathit{A}} \sum_{\beta \in \mathit{A}} P_{\alpha \beta}(d) \log_{2}\frac{P_{\alpha \beta}(d)}{P_{\alpha}P_{\beta}},
\label{mutualinfo}
\end{equation}
for symbols $\alpha,\beta\in\mathit{A}$ (in the case of DNA, $\mathit{A}=\left\{a,t,c,g\right\}$). $P_{\alpha\beta}(d)$ is the probability that symbols $\alpha$ and $\beta$ are found a distance $d$ apart. This is related to the correlation function in Eq.~\ref{correlation}~\cite{mutualinfo}:
\begin{equation}
\Gamma (d) = \displaystyle  \sum_{\alpha \in \mathit{A}} \sum_{\beta \in \mathit{A}} a_{\alpha} a_{\beta}P_{\alpha\beta} (d)-\left(\sum_{\alpha \in \mathit{A}} a_{\alpha}P_{\alpha}\right)^{2},
\label{correlation}
\end{equation}
where $a_{\alpha}$ and $a_{\beta}$ are numerical representations of symbols $\alpha$ and $\beta$. As discussed by Li~\cite{mutualinfo}, the fact that we are working with a finite sequence means that this $M(d)$ overestimates the true $M_{\scriptscriptstyle T}(d)$ by
\begin{equation}
M(d)-M_{\scriptscriptstyle T}(d)\approx \frac{K\left(K-2\right)}{2N},
\label{overestimate}
\end{equation}
where $K$ is the number of symbols (for DNA this is always 4) and $N$ is the sequence length. The shortest sequence used was the sequence of the {\it Homo sapiens} immunoglobulin superfamily, member 8 gene (GenBank accession BC004108), which was $N=1750$ base pairs in length. Thus for this gene the difference between the estimated and real mutual information is  $\approx \frac{4\times2}{2\times 1750} = 0.002$, which is an order of a factor of ten less than the mutual information estimate for this gene. Furthermore, since in our results below we compare the mutual information of the sequence with that of the randomized sequence, we are effectively eliminating this inaccuracy. 

The mutual information is (at least for large $d$) proportional to the correlation squared, $\Gamma(d)$~\cite{mutualinfo}. Even for small $d$, the mutual information function is still providing an estimate of the correlations. The range of $d$ we used (up to 1024) means we are providing a reasonable estimate of the correlations at these larger distances. In biological terms, we are capturing correlations within regions of genes, and between promoter regions and DNA. This length is not sufficiently large to explore longer range correlations such as those between genes (typically tens of thousands of bases) or those that might exist between activator or silencer regions and promoters, again on the order of tens of thousands of bases~\cite{LevineTjian}. In the whole chromosome analysis we are finding repeating elements and other correlations in junk DNA in addition to correlations within genes.
\subsection{Higuchi fractal measure}
A method for determining correlations in sequences is to use the Higuchi fractal method~\cite{Higuchi}.
In using this method we compute
\begin{equation}
L\left(k\right)=\displaystyle\sum_{m=0}^{k-1}\frac{N-1}{\lfloor\frac{N-m}{k}\rfloor k^{2}}
\displaystyle\sum_{i=1}^{\lfloor\frac{N-m}{k}\rfloor}\abs\left(x\left(m+ik\right)-x\left(m+\left(i-1\right)k\right)\right),
\label{hfractal}
\end{equation}
for $k=1,\ldots,1024$ over non-overlapping subsequences of length 4000. The sequence $x\left(i\right)$ is generated by mapping the sequence of bases, $s\left(i\right)$:
\begin{equation}
x\left(i\right)=
\begin{cases}
~~1.0, & s(i)=a,\\
~~0.5, & s(i)=t,\\
-0.5, & s(i)=c,\\
-1.0, & s(i)=g.\\
\end{cases}
\label{hmap}
\end{equation}
Performing linear regression on $\log L\left(k\right)$ versus $\log k$ then gives a slope of $-D$, where $D$ is the estimate of the true fractal measure. For a high degree of correlation, we expect a value of $D$ closer to one.

One can also apply the Higuchi method to the density of bases in blocks, as carried by Lu {\it et al.}~\cite{Lu}, however this does not provide a measure of correlations in the sequence as the authors claim, but rather correlations in the density function. In the fashion we use it, we are detecting correlations in the sequence, though as with the mutual information function we only explore correlations up to 1024 base pairs.
\section{Results}
\subsection{Short DNA sequences}
\begin{figure}[ht]
 \centering\mbox{
        	\subfigure[This figure shows the plot of the mutual information function $M(d)$ in Eq.~\ref{mutualinfo} against base distance $d$ for the sequence of the MAP kinase-activated protein kinase 2 gene from {\it Mus musculus}, shown in a darker line style, compared with the set of 100 randomized sequences of the same base distribution, the lighter band. The graph of mutual information in the MAP kinase gene mostly sits about the ``noise floor'' of the randomized sequences, in which the correlations have been destroyed.]{
            \label{miplots:real}
            \includegraphics[width=6cm]{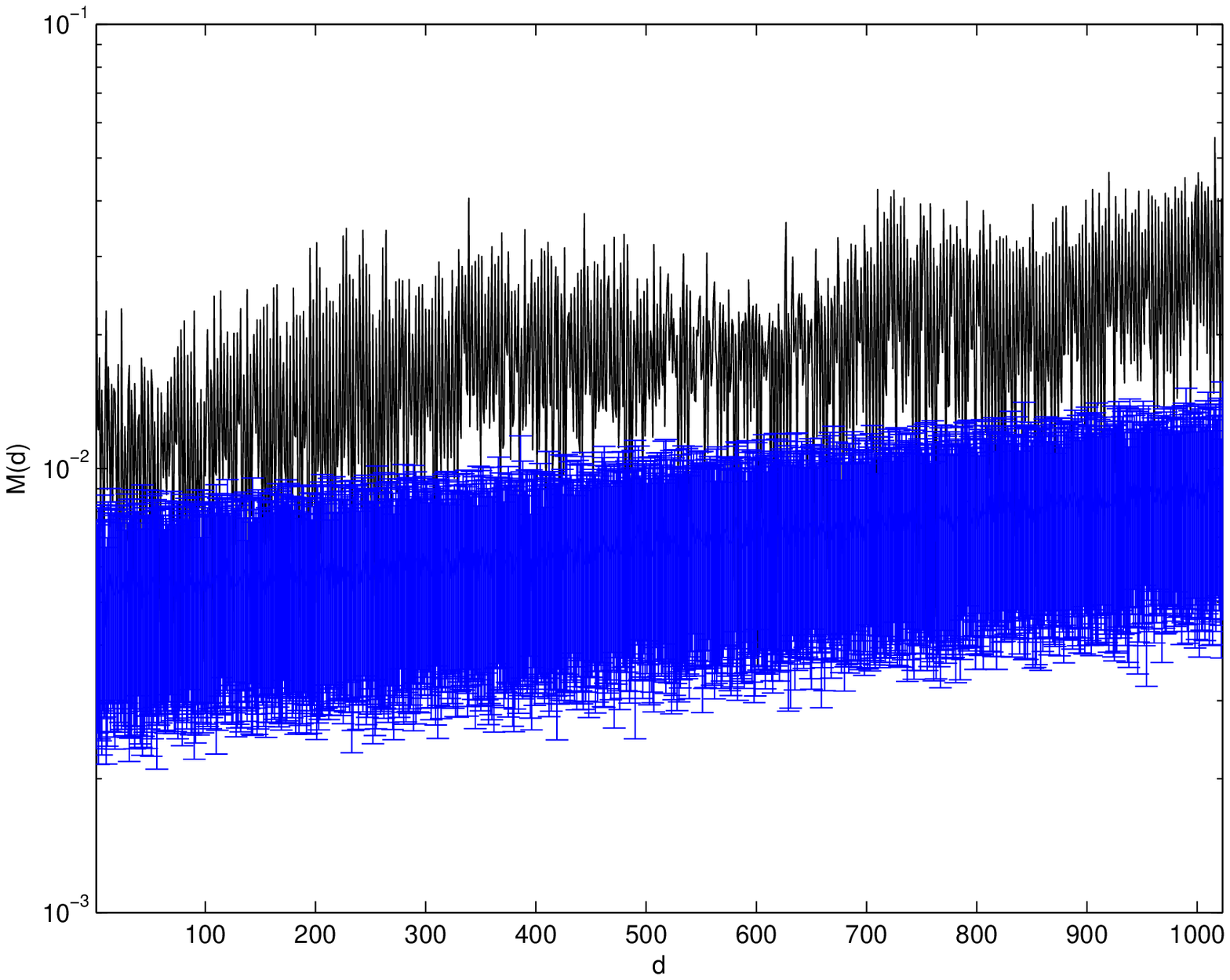}
        }}\mbox{
        \subfigure[This figure shows the plot of the mutual information function $M(d)$ in Eq.~\ref{mutualinfo} against base distance $d$ for the virtual DNA sequence number $14$, shown in a darker line style, compared with the set of 100 randomized sequences of the same base distribution, shown as a lighter band. The graphs mostly overlap, indicating few significant correlations in the virtual sequence when compared with the randomized sequences containing little to no correlations.]{
            \label{miplots:virtual}
            \includegraphics[width=6cm]{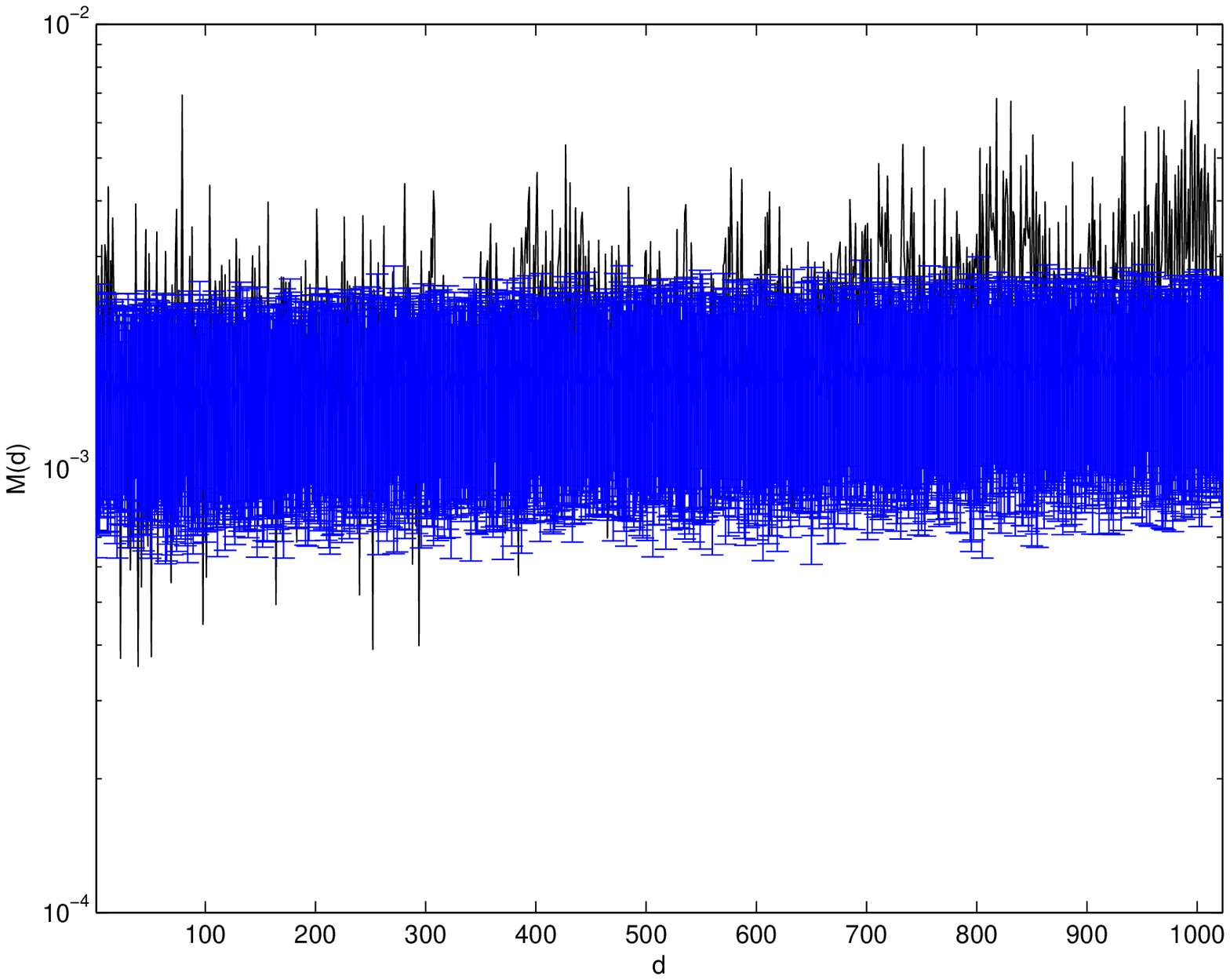}
        }}
        \caption{These figures show the plots of the mutual information function $M(d)$ in against base distance $d$ for (a) a real sequence and (b) a virtual sequence. At larger distances, there are fewer symbols at that distance that are available for computing the mutual information, so the over-estimates increases in value, producing a slight slope to the graphs.}
\label{miplots}
\end{figure}

To analyze the short DNA sequences (real, virtual, and random) using the mutual information function~\ref{mutualinfo}, we compared the mutual information plot with the average +/- standard deviation plot of the mutual information function for 100 randomized sequences with the same base distribution but in random order (thus eliminating correlations). Examples of this are shown in Fig.~\ref{miplots}. 

We determined the maximum distance at which significant correlations were present, up to the maximum distance studied of 1024. The results of this for the 20 real, virtual, and random sequences are shown in Table~\ref{dists}. No long-range correlations are present in our benchmark random sequences as one would expect, however correlations up to distance $d>1024$ are present in our virtual sequences, and even longer range correlations of distance $d>1024$ can be found in real sequences. Because the mutation process used to generate the virtual sequences was random, there was a significant variation in the length of correlations present. This corresponded well to the number of repeated elements and copy mutations, in particular with the copy mutations. Future work will attempt to quantify the mutual information values with a directed model of evolution where we take real sequences and apply mutation operators in a realistic fashion, for example point mutations are much more likely to be seen in the ``wobble'' positions of codons than elsewhere, and this in turn is much more likely than insertions and deletions.
\begin{table}[ht]
\centering
\caption{This table shows the approximate ($\pm 50$) distances at which the mutual information function drops down to the level of the uncorrelated sequences of the same base distribution. The numbering of the real sequences matches the ordering they are given in Table~\ref{real}. The numbering of the virtual sequences corresponds to the random sequence which was mutated to produce that virtual sequence, but bears no relationship to the numbering of the real sequences.}
\begin{tabular}{|c|c|c|c|}\hline
Sequence number & Random & Virtual & Real\\\hline
1 & 0 & 0 & $>1024$\\\hline
2 & 0 & 0 & $>1024$\\\hline
3 & 0 & $>1024$ & $700$\\\hline
4 & 0 & 100 & $800$\\\hline
5 & 0 & 50 & $0$\\\hline
6 & 0 & 0 & $>1024$\\\hline
7 & 0 & 850 & $>1024$\\\hline
8 & 0 & 0 & $>1024$\\\hline
9 & 0 & 0 & $>1024$\\\hline
10 & 0 & 800 & $>1024$\\\hline
11 & 0 & 0 & $>1024$\\\hline
12 & 0 & $>1024$ & $>1024$\\\hline
13 & 0 & 100 & $950$\\\hline
14 & 0 & $>1024$ & $600$\\\hline
15 & 0 & $>1024$ & $>1024$\\\hline
16 & 0 & 0 & $>1024$\\\hline
17 & 0 & 0 & $>1024$\\\hline
18 & 0 & 0 &$>1024$ \\\hline
19 & 0 & $>1024$ & $>1024$\\\hline
20 & 0 & 0 & $>1024$\\\hline
\end{tabular}
\label{dists}
\end{table}

The results of using the Higuchi fractal method are shown in Table~\ref{htable}. Note that these estimates are relatively independent of the choice of mapping of bases onto numbers (several different mappings were tried with variations on the order of $0.001$), and the numbers are in fact overestimates of the true fractal dimension. The fractal dimensions appear unrelated to the mutual information distances, thus illustrating the fact that the mutual information function is a better characterization of the distances at which correlations are present. 
\begin{table}[ht]
\centering
\caption{This table shows the estimates of the fractal dimension as ascertained using the Higuchi method described by Eq.~\ref{hfractal}. The numbering of the real sequences matches the ordering they are given in Table~\ref{real}. The numbering of the virtual sequences corresponds to the random sequence which was mutated, but bears no relationship to the numbering of the real sequences.}
\begin{tabular}{|c|c|c|c|}
\hline
Sequence number & Random & Virtual & Real\\\hline
1 & 1.104 & 1.103 & 1.098\\\hline
2 & 1.103 & 1.094 & 1.095\\\hline
3 & 1.104 & 1.094 & 1.118\\\hline
4 & 1.104 & 1.086 & 1.110\\\hline
5 & 1.103 & 1.086 & 1.092\\\hline
6 & 1.102 & 1.094 & 1.103\\\hline
7 & 1.105 & 1.100 & 1.105\\\hline
8 & 1.103 & 1.102 & 1.087\\\hline
9 & 1.102 & 1.093 & 1.080\\\hline
10 & 1.103 & 1.099 & 1.099\\\hline
11 & 1.103 & 1.089 & 1.087\\\hline
12 & 1.104 & 1.103 & 1.098\\\hline
13 & 1.103 & 1.099 & 1.098\\\hline
14 & 1.104 & 1.091 & 1.055\\\hline
15 & 1.104 & 1.100 & 1.101\\\hline
16 & 1.103 & 1.103 & 1.090\\\hline
17 & 1.102 & 1.102 & 1.097\\\hline
18 & 1.102 & 1.091 & 1.094\\\hline
19 & 1.102 & 1.099 & 1.099\\\hline
20 & 1.103 & 1.099 & 1.091\\\hline
\end{tabular}
\label{htable}
\end{table}
\subsection{Whole chromosome sequences}
The results of analyzing chromosomes from {\it E.~coli}, {\it M.~musculus}, and {\it H. sapiens} using both the Higuchi fractal measure, $D$, and the mutual information function, $M(d)$, indicate the presence of correlations up to the maximum length explored ($1024$). This is shown in Table~\ref{reswg}. There is less variation in these measures for {\it E.~coli}, which has a greater proportion of gene-coding DNA to other sequences, these gene-coding regions allow less room for repeating elements due to evolutionary and size constraints, and thus have a lower correlation distance.
\begin{table}[ht]
\centering
\caption{This table shows the average Higuchi fractal dimension $D$ over blocks of length $4000$ in the chromosomes listed, along with the variance, and the distance $d$ at which correlations exist as determined by mutual information function in Eq.~\ref{mutualinfo}.}
\begin{tabular}{|c|c|c|c|}
\hline
Sequence & $\mathrm{mean}\left(D\right)$ & $\mathrm{var}\left(D\right)$ & $d$\\\hline
{\it Eschercia coli} K12, complete genome & 1.10039 & $2.07\times10^{-5}$ & $>1024$ \\\hline
{\it Mus musculus} chromosome 2 & 1.09691 & $7.59\times10^{-5}$ & $>1024$\\\hline
{\it Homo sapiens} chromosome 20 & 1.089 & 0.00991 & $>1024$\\\hline
\end{tabular}
\label{reswg}
\end{table}
\section{Conclusions}
We found long-range correlations present in short sequences of real DNA, ``virtual'' DNA, and throughout whole chromosomes. Our simulation of genetic mutation events in ``junk'' DNA with fill, copy, and mutate operations also produces long range-correlations approaching 1024 bases in length. Our negative test, with computer generated random sequences, succeeds in that we do not find any significant long-range correlations. These results confirm that mutational events in non-conserved regions of DNA can give rise to long-range correlations. 
\section*{Acknowledgements}
We acknowledge the funding provided by The University of Adelaide. Useful discussions with Wentian Li are greatly appreciated.

\end{document}